\documentclass{article}
\usepackage{graphicx}
\usepackage{amstext,amssymb,amsbsy}
\usepackage{cite}

\author{
   Niels G. Gresnigt \and Adam B. Gillard\\
}

\title{Magnetic moments of octet baryons with deformed $U_q(su(3))$ flavor symmetry}
\begin{document}
\maketitle
\begin{abstract}
The effect of a $q$-deformation of flavor symmetry on octet baryon magnetic moments is investigated by taking the quantum group $U_q(su(3))$ as a flavor symmetry and calculating expressions for the magnetic moments of the octet baryons. Using the experimental values for the magnetic moments of the proton, neutron, and $\Lambda$ baryons as input, the magnetic moments of the up, down, and strange quarks, and by extension those of the remaining octet baryons, depend on the value of the deformation parameter $q$. Plotting the least square error as a function of $q$ we find the undeformed value $q=1$ is a local maximum. Nearby are a relative minimum (at $q=0.610$) and an absolute minimum (at $q=1.774$). The value $q=1.774$ gives an approximately 10\% decrease in the least square error compared to the undeformed case. For this value of $q$ the quark masses of the up and down quarks are calculated to be 331 MeV and 305 MeV respectively.
\end{abstract}

\section{Introduction}
We investigate how the magnetic moments of the octet baryons are affected when the classical flavor group $SU(3)$ is replaced by its quantum group counterpart $U_q(su(3))$. Such a quantum group represents a $q$-deformation of the universal enveloping algebra $U(su(3))$ leading to a generalization of the familiar symmetry concepts. 

Even though the magnetic moments of nucleons and other octet baryons as calculated within the non-relativistic quark model based on $SU(3)$ flavor symmetry are in reasonable agreement with experimental values, a first principle derivation of nucleon magnetic moments is still lacking at present. Taken as a flavor symmetry, the quantum group $U_q(su(3))$ gives rise to exceptionally accurate baryon mass formulas \cite{gavrilik1997quantum,gavrilik2001quantum} that agree perfectly with the experimental data, especially when the electromagnetic contributions to mass are accounted for \cite{Gresnigt2016baryons}. Given this exceptional accuracy of baryon mass formulas that result from a deformation of flavor symmetry, it is  natural and worthwhile to likewise investigate the effect of such a deformation on the magnetic moments of baryons. In the present paper we restrict ourselves to octet baryons and derive expressions for the octet baryons magnetic moments in terms of light quark magnetic moments based on a $q$-deformed flavor symmetry. 

Quantum groups (which are algebras rather than groups) provide a generalization of familiar symmetry concepts through the deformation of Lie groups and algebras. These deformations of classical Lie groups (algebras) into quantum groups (algebras) extends the domain of classical group theory. Such quantum groups are deformations on Hopf algebras and depend on a deformation parameter $q$ with the value $q=1$ returning the undeformed universal enveloping algebra. First formalized by Jimbo \cite{jimbo1985aq} and Drinfeld \cite{drinfeld1985soviet} as a class of Hopf algebras, quantum groups are closely realted to noncommutative geometry as well as certain invariants of knots. As such, they have found many applications in theoretical physics, see \cite{finkelstein2001q,Finkelstein2012,Steinacker1998,Sternheimer2007,Majid1994,Lukierski2003,castellani1996quantum} and references therein, and there is evidence from quantum gravity that spacetime symmetries should be $q$-deformed \cite{bianchi2011note}.

A good overview of the calculation of light baryon magnetic moments in the non-relativistic quark model is given in chapter 9 of \cite{thomson2013modern}. The wavefunction for a bound 3-quark state, accounting for all degrees of freedom can be written as
\begin{eqnarray}
\psi=\phi_{\textrm{flavor}}\chi_{\textrm{spin}}\xi_{\textrm{color}}\eta_{\textrm{space}}.
\end{eqnarray}
The fermionic nature of quarks means that this overall wavefunction must be antisymmetric under the interchange of any two quarks. As a result, the flavor-spin wavefunction $\phi_{\textrm{flavor}}\chi_{\textrm{spin}}$ for $L=0$ baryons must be symmetric. The combination
\begin{eqnarray}
\psi=\frac{1}{\sqrt{2}}\left( \phi_S\chi_S+\phi_A\chi_A\right) 
\end{eqnarray}
consisting of mixed-symmetric and mixed-antisymmetric flavor states $\phi_S$ and $\phi_A$ and the mixed-symmetric and mixed-antisymmetric spin states $\chi_S$ and $\chi_A$ is symmetric under interchange of any two quarks. 

One result of deforming the flavor space is that the classical notion of a fermion is lost. This is so because the underlying space is now a braided space and the statistics are those of the fermionic quantum plane. The fermion flavor state is now an eigenstate of the braid operator rather than the permutation operator with the braid operator dictating how two quarks may be interchanged. The classical notion of a fermion may be restored by considering additional deformations in the remaining spaces (for example spin) but we defer such an in depth study involving multiple deformations for a future occassion. 

We write the two possible spin-flavor states corresponding to the proton (uud) and the neutron (ddu) as
\begin{eqnarray}
\vert p\uparrow\rangle &=&\frac{1}{\sqrt{2}}\left[  \phi_S\left( \frac{1}{2},+\frac{1}{2}\right)  \chi_S\left( \frac{1}{2},+\frac{1}{2}\right)+\phi_A\left( \frac{1}{2},+\frac{1}{2}\right)\chi_A\left( \frac{1}{2},+\frac{1}{2}\right)\right] ,\\
 \vert n\uparrow\rangle &=&\frac{1}{\sqrt{2}}\left[  \phi_S\left( \frac{1}{2},-\frac{1}{2}\right)  \chi_S\left( \frac{1}{2},+\frac{1}{2}\right)+\phi_A\left( \frac{1}{2},-\frac{1}{2}\right)\chi_A\left( \frac{1}{2},+\frac{1}{2}\right)\right].
\end{eqnarray}
The magnetic moment of the proton is written as
\begin{eqnarray}
\mu_p=\langle \hat{\mu}_z\rangle =\langle p\uparrow\vert \hat{\mu}_z^{(1)}+\hat{\mu}_z^{(2)}+\hat{\mu}_z^{(3)}\vert p\uparrow\rangle,
\end{eqnarray}
evaluated using
\begin{eqnarray}
\hat{\mu}_z\vert u\uparrow\rangle &=&+\mu_u\vert u\uparrow\rangle,\qquad \hat{\mu}_z\vert u\downarrow\rangle=-\mu_u\vert u\downarrow\rangle,\\
\hat{\mu}_z\vert d\uparrow\rangle &=&+\mu_d\vert d\uparrow\rangle,\qquad \hat{\mu}_z\vert d\downarrow\rangle =-\mu_d\vert d\downarrow\rangle,
\end{eqnarray}
to obtain
\begin{eqnarray}
\mu_p=\frac{4}{3}\mu_u-\frac{1}{3}\mu_d.
\end{eqnarray}
Similarly, the magnetic moment of the neutron may be calculated. Calculation of the remaining octet baryons containing a strange quark (hyperons) requires making use of the full $SU(3)$ group structure and the ladder operators corresponding to the simple roots. This general process carries over to the deformed case although now the flavor states and ladder operators are all modified.

In section \ref{quantumgroups} we introduce the necessary mathematics relating to quantum groups. In particular we give an outline of the quantum group $U_q(su(3))$ and define the ladder operators along with their coproducts. In section \ref{flavorstates} we use the $q$-deformed ladder operators to find all the $U_q(su(3))$ flavor states up to total isospin $I=3/2$. From these we derive the isospin $I=1/2$ $q$-symmetric and $q$-antisymmetric flavor states. Combining these flavor states with undeformed spin states, in section \ref{protonneutron} we write down the deformed state of a spin-up proton before using this calculating calculate the magnetic moment of the proton and neutron, and subsequently quark masses in section \ref{quarkmasses}. To obtain an expression for the strange quark in terms of the deformation paramtere $q$, in section \ref{lambdastate} we calculate the magnetic moment of the $\Lambda$ baryon. In section \ref{analysis} we compare our results to experimental data and find that for certain values of (real) $q$, the deformed theory agrees better with experimental data.

\section{Quantum groups and algebras}\label{quantumgroups}
Only a very brief introduction to the relevant aspects of quantum groups is provided here. The literature on quantum groups and algebras is extensive. For an excellent introduction the reader is directed to \cite{jaganathan2001some}. A more complete text is \cite{majid2000foundations}. Various applications of quantum groups to physics are discussed in \cite{castellani1996quantum}.

The quantum (enveloping) algebra $U_q(su(n))$ corresponding to a one-parameter deformation of the universal enveloping algebra of $su(n)$, is a Hopf algebra with unit $\mathbf{1}$ and generators $H_i$, $X_i^{\pm}$, $i=1,2,...,n-1$, defined through the commutation relations in the Cartan-Chevalley basis as 
\begin{eqnarray}
\left[ H_i, H_j\right] &=&0\\
\left[ H_i, X_j^{\pm}\right] &=&a_{ij}X_j^{\pm}\\
\left[ X_i^+, X_j^-\right]&=&\delta_{ij}[H_i]_q\equiv\delta_{ij}\frac{q^{H_i/2}-q^{-H_i/2}}{q^{1/2}-q^{-1/2}}\label{quantalg3},
\end{eqnarray}
together with the quadratic and cubic deformed $q$-Serre relations
\begin{eqnarray}
\left[ X_i^{\pm}, X_j^{\pm}\right]=0,\;\; j\neq i\pm 1,\;\; 1\leq i,j\leq n-1
\end{eqnarray}
and
\begin{eqnarray}\label{cubicserre}
(X_i^{\pm})^2 X_j^{\pm}-[2]_q X_i^{\pm}X_j^{\pm}X_i^{\pm}+X_j^{\pm}(X_i^{\pm})^2=0,\;\; j=i\pm 1,\;\; 1\leq i,j\leq n-1
\end{eqnarray}
respectively \cite{jimbo1985aq,quesne1992complementarity}. Here $a_{ij}$ is an element of the Cartan matrix
 \begin{displaymath}
   a_{ij} = \left\{
     \begin{array}{lr}
       2 & j=i\\
       -1 & j=i\pm 1\\
       0 & \textrm{otherwise}.
     \end{array}
   \right.
\end{displaymath} 
The $q$-number
\begin{eqnarray}
[N]_q=\frac{q^{N/2}-q^{-N/2}}{q^{1/2}-q^{-1/2}}
\end{eqnarray}
is defined for both operators (as in equation (\ref{quantalg3})) and real numbers\footnote{In this paper we will only have to deal with integer values of $N$. The definition however holds for real numbers.} (as in equation (\ref{cubicserre})). In particular, $[2]_q=q+q^{-1}$. The definition of the algebra is completed by the Hermiticity properties
\begin{eqnarray}
(H_i)^{\dagger}=H_i,\qquad (X_i^{\pm})^{\dagger}=X_i^{\mp}.
\end{eqnarray}

Restricting ourselves to $SUq(2)$, the actions of the ladder operators on an isospin state are defined as
\begin{eqnarray}\label{coproductT}
\hat{T}_{\pm}\vert I, I_3\rangle=\sqrt{[I\mp I_3]_q[I\pm I_3+1]_q}\vert I, I_3\pm 1\rangle
\end{eqnarray}
Where $I$ and $I_3$ are the total isospin and third component of isospin respectively. The coproducts associated wit the ladder operators are
\begin{eqnarray}
\Delta\hat{T}_z=1\otimes \hat{T}_z+\hat{T}_z\otimes 1,\qquad \Delta\hat{T}_{\pm}=q^{-\frac{\hat{T}_3}{2}}\otimes \hat{T}_{\pm}+\hat{T}_{\pm}\otimes q^{\frac{\hat{T}_3}{2}},
\end{eqnarray}
where $T_3$ is the third component of spin. Using the co-associativity of the coproduct we determine the coproduct $(\Delta\otimes\mathrm{id})\Delta\hat{T}_\pm=(\mathrm{id}\otimes\Delta)\Delta\hat{T}_\pm$ to be
\begin{eqnarray}\label{coassociativity}
(\Delta\otimes\mathrm{id})\Delta\hat{T}_\pm=q^{-\frac{\hat{T}_z}{2}}\otimes q^{-\frac{\hat{T}_z}{2}}\otimes\hat{T}_\pm+q^{-\frac{\hat{T}_z}{2}}\otimes \hat{T}_\pm\otimes q^{\frac{\hat{T}_z}{2}}+\hat{T}_\pm\otimes q^{\frac{\hat{T}_z}{2}}\otimes q^{\frac{\hat{T}_z}{2}}.
\end{eqnarray}

To include strange quarks, we need our braided flavor space to have $q$-isospin symmetry embedded into $U_q(su(3))$ via the Gelfand-Tsetlin reduction chain
\begin{eqnarray}
U_q(u_{\hat{T}_z}(1))\subset U_q(u_{\hat{Y}}(1))\otimes U_q(su_{\hat{T}}(2))\subset U_q(su(3))
\end{eqnarray}
where $\hat{Y}$ is the hypercharge operator. The quantum group $U_q(su(3))$ has rank two so there are two simple roots. Labeling the basis states as $u\equiv\left|\frac{1}{2},\frac{1}{3}\right\rangle$, $d\equiv\left|-\frac{1}{2},\frac{1}{3}\right\rangle$ and $s\equiv\left|0,-\frac{2}{3}\right\rangle$ where the first entry is the isospin label and the second is the hypercharge, we have the following relations
\begin{eqnarray}
\nonumber\quad \hat{T}_+d=u,\quad \hat{T}_-u=d, \quad \hat{U}_+s=d,\quad\hat{U}_-d=s,\quad\hat{V}_+s=u,\quad\hat{V}_-u=s,\\
\nonumber\hat{T}_\pm\left|t_z,y\right\rangle\propto\left|t_z\pm1,y\right\rangle,\;\hat{U}_\pm\left|t_z,y\right\rangle\propto \left|t_z\mp\frac{1}{2},y\pm1\right\rangle,\;\hat{V}_\pm\left|t_z,y\right\rangle\propto\left|t_z\pm\frac{1}{2},y\pm1\right\rangle,
\end{eqnarray}
and
\begin{eqnarray}
\nonumber &&q^{\pm\frac{\hat{T}_z}{2}}u=q^{\pm\frac{1}{2}}u,\quad q^{\pm\frac{\hat{T}_z}{2}}d=q^{\pm\frac{1}{2}}d,\quad q^{\pm\frac{\hat{T}_z}{2}}s=s,\quad q^{\pm\frac{\hat{U}_0}{2}}u=u,\quad q^{\pm\frac{\hat{U}_0}{2}}d=q^{\pm\frac{1}{2}}d,\\
\nonumber &&q^{\pm\frac{\hat{U}_0}{2}}s=q^{\mp\frac{1}{2}}s,\quad q^{\pm\frac{\hat{V}_0}{2}}u=q^{\pm\frac{1}{2}}u,\quad q^{\pm\frac{\hat{V}_0}{2}}d=d,\quad q^{\pm\frac{\hat{V}_0}{2}}s=q^{\mp\frac{1}{2}}s.
\end{eqnarray}
The $q$-isospin coproducts $\Delta\hat{T}_\pm$ are as before. The coproduct corresponding to the second simple root is
\begin{eqnarray}
\Delta\hat{U}_\pm=q^{-\frac{\hat{U}_0}{2}}\otimes\hat{U}_\pm+\hat{U}_\pm\otimes q^{\frac{\hat{U}_0}{2}}
\end{eqnarray}
for completeness we include the coproduct of the non-simple root although we will not make use of it explicitly.
\begin{eqnarray}
\Delta\hat{V}_+&=&q^{-\frac{\hat{V}_0}{2}}\otimes\hat{V}_++\hat{V}_+\otimes q^{\frac{\hat{V}_0}{2}}+(1-q^{-2})q^{-\frac{\hat{T}_z}{2}}\hat{U}_+\otimes\hat{T}_+q^{\frac{\hat{U}_0}{2}}\\
\Delta\hat{V}_-&=&q^{-\frac{\hat{V}_0}{2}}\otimes\hat{V}_-+\hat{V}_-\otimes q^{\frac{\hat{V}_0}{2}}+(1-q^2)q^{-\frac{\hat{U}_0}{2}}\hat{T}_-\otimes\hat{U}_-q^{\frac{\hat{T}_z}{2}}.
\end{eqnarray}
\subsection{Braided statistics}\label{braidstatistics}

The fundamental basis states being acted on are the isospin-up up quark flavour state $u\equiv\left|\frac{1}{2},\frac{1}{2}\right\rangle$ and the isospin-down down quark flavour state $d\equiv\left|\frac{1}{2},\frac{-1}{2}\right\rangle$. In the undeformed case fermion flavour states are eigenstates of the permutation operator $P$, which for two identical particles $\left|a\right\rangle$ and $\left|b\right\rangle$ is defined as\footnote{the physical fermions have additional defining aspects such at spin, colour and space degrees of freedom and it is the overall state must be completely antisymmetric.} $P_{12}\left|a\right\rangle\otimes\left|b\right\rangle=\left|b\right\rangle\otimes\left|a\right\rangle$, $P_{12}^2=1$. We have a four-dimensional space spanned by the permutation operator eigenstates $\left|a\right\rangle\otimes\left|a\right\rangle$, $\frac{1}{\sqrt{2}}[\left|a\right\rangle\otimes\left|b\right\rangle+\left|b\right\rangle\otimes\left|a\right\rangle]$, $\left|b\right\rangle\otimes\left|b\right\rangle$ with eigenvalue $+1$ and $\frac{1}{\sqrt{2}}[\left|a\right\rangle\otimes\left|b\right\rangle-\left|b\right\rangle\otimes\left|a\right\rangle]$ with eigenvalue $-1$. In matrix notation the permutation operator takes the form 
\begin{eqnarray}
P=\left(\begin{array}{cccc}
1&0&0&0 \\
0&0&1&0 \\
0&1&0&0 \\
0&0&0&1
\end{array}\right).
\end{eqnarray}
In the deformed setting the notion of fermion flavor states needs modification. They must now be eigenstates of the braid operator $B\equiv PR$, where 
\begin{eqnarray}
R=\left(\begin{array}{cccc}
1&0&0&0 \\
0&q^{\frac{1}{2}}&0&0 \\
0&1-q&q^{\frac{1}{2}}&0 \\
0&0&0&1
\end{array}\right),
\end{eqnarray}
is the universal $R$-matrix\footnote{See \cite{jaganathan2001some} for an introductory discussion on the $R$-matrix.}. 

From now on we supress the tensor product and write $uu\equiv u\otimes u$ etc. The normalized states of the braid operator are $uu$, $\frac{1}{\sqrt{[2]_q}}[q^{\frac{-1}{4}}ud+q^{\frac{1}{4}}du]$ and $dd$ with eigenvalues $+1$ and $\frac{1}{\sqrt{[2]_q}}[-q^{\frac{1}{4}}ud+q^{\frac{-1}{4}}du]$ with eigenvalue $-q$. The first three states define the notion of $q$-symmetric while the last state defines the notion of $q$-antisymmetric. Forming the vector $X^t=(uu,ud,du,dd)$, one obtains the braided interchange relations of the elements of this vector by considering $BX=X'$,
\begin{equation}\label{quarkstatistics}
B(uu)=uu,\quad B(ud)=(1-q)ud+q^{\frac{1}{2}}du,\quad B(du)=q^{\frac{1}{2}}ud,\quad B(dd)=dd.
\end{equation}
We will also need the corresponding expressions resulting from introducing the strange quark $s$,
\begin{eqnarray}
B(uu)=uu,\quad B(us)=(1-q)us+q^{\frac{1}{2}}su,\quad B(su)=q^{\frac{1}{2}}us,\quad B(ss)=ss \\
B(dd)=dd,\quad B(ds)=(1-q)ds+q^{\frac{1}{2}}sd,\quad B(sd)=q^{\frac{1}{2}}ds,\quad B(ss)=ss.
\end{eqnarray}

\section{$U_q(su(2))$ flavor states}\label{flavorstates}

Starting with a total isospin 1 state $\phi(1,-1)=dd$ and applying the ladder operator $\hat{T}_+$ to the left hand side and equivalently applying the coproduct $\Delta\hat{T}_+$ to the right hand side to obtain the other two isospin 1 states

\begin{eqnarray}
\hat{T}_+\phi(1,-1)&=&\sqrt{[2]_q}\phi(1,0),\\
\Delta\hat{T}_+(dd)&=&q^{1/4}d\otimes\hat{T}_+(d)+\hat{T}_+(d)\otimes q^{-1/4}d,\\
&=& q^{1/4}du+q^{-1/4}ud.
\end{eqnarray}
We therefore obtain
\begin{eqnarray}
\phi(-1,0)=\frac{1}{\sqrt{[2]_q}}\left( q^{1/4}du+q^{-1/4}ud\right). 
\end{eqnarray}

For an isospin $3/2$ state, for example $\phi\left( \frac{3}{2},-\frac{3}{2}\right)=ddd$, one must use the co-associativity of the coproduct (as expressed in equations (\ref{coassociativity})) when evaluation the action of the ladder operator of the right hand side. Applying the ladder operator to both sides we obtain
\begin{eqnarray}
\hat{T}_+\phi\left(\frac{3}{2},-\frac{3}{2}\right) &=& \sqrt{[3]_q}\phi\left( \frac{3}{2},-\frac{1}{2}\right),\\
\nonumber(\Delta\otimes \textrm{id})\Delta\hat{T}_+(ddd)&=&q^{-\frac{\hat{T}_z}{2}}(d)\otimes q^{\frac{-\hat{T}_z}{2}}(d)\otimes\hat{T}_\pm(d)+q^{-\frac{\hat{T}_z}{2}}(d)\otimes \hat{T}_\pm(d)\otimes q^{\frac{\hat{T}_z}{2}}(d)\\
\nonumber&+&\hat{T}_\pm(d)\otimes q^{\frac{\hat{T}_z}{2}}(d)\otimes q^{\frac{\hat{T}_z}{2}}(d)\\
\nonumber&=& q^{1/2}ddu+dud+q^{-1/2}udd.
\end{eqnarray}
so that
\begin{eqnarray}
\phi\left(\frac{3}{2},-\frac{1}{2}\right)=\frac{1}{\sqrt{[3]_q}}\left( q^{1/2}ddu+dud+q^{-1/2}udd\right). 
\end{eqnarray}
Continuing this process one obtain all the isospin 1 and isospin and 3/2 (included for completeness) states

\begin{eqnarray}
\phi(-1,1)&=&dd,\\
\phi(1,0)&=&\frac{1}{\sqrt{[2]_q}}(q^{1/4}du+q^{-1/4}ud),\\
\phi(1,1)&=&uu.
\end{eqnarray}
and the orthogonal singlet state
\begin{eqnarray}
\phi(0,0)=\frac{1}{\sqrt{[2]_q}}(q^{-1/4}du-q^{1/4}ud).
\end{eqnarray}
For isospin 3/2 one has
\begin{eqnarray}
\phi\left( \frac{3}{2},\frac{-3}{2}\right) &=&ddd,\\
\phi\left( \frac{3}{2},\frac{-1}{2}\right) &=&\frac{1}{\sqrt{[3]_q}}(q^{1/2}ddu+dud+q^{-1/2}udd),\\
\phi\left( \frac{3}{2},\frac{1}{2}\right) &=&\frac{1}{\sqrt{[3]_q}}(q^{1/2}duu+udu+q^{-1/2}uud),\\
\phi\left( \frac{3}{2},\frac{-3}{2}\right) &=&uuu.
\end{eqnarray}
These can be used to construct the flavor states of the decuplet baryons although we will not do so in the present paper.

The mixed $q$-symmetric and $q$-antisymmetric (in the sense discussed in subsection \ref{braidstatistics}) states are
\begin{eqnarray}
\phi_S\left(\frac{1}{2},\frac{-1}{2}\right)&=&\frac{-q^{\frac{1}{4}}}{\sqrt{[2]_q [3]_q}}\left( [2]_q q^{-1/2}ddu-udd-q^{1/2}dud\right),\\
\phi_S\left(\frac{1}{2},\frac{1}{2}\right)&=&\frac{q^{\frac{-1}{4}}}{\sqrt{[2]_q [3]_q}}\left( [2]_q q^{1/2}uud-duu-q^{-1/2}udu\right),
\end{eqnarray}
and
\begin{eqnarray}
\phi_A\left(\frac{1}{2},\frac{-1}{2}\right)&=&\frac{1}{\sqrt{[2]_q}}\left(q^{-1/4}dud-q^{1/4}udd\right),\\
\phi_A\left(\frac{1}{2},\frac{1}{2}\right)&=&\frac{1}{\sqrt{[2]_q}}\left( -q^{1/4}udu+q^{-1/4}duu\right),
\end{eqnarray}
where by $q$-symmetry we mean that the exchange of two quarks satisfies equation (\ref{quarkstatistics}) with eigenvalue 1 and by $q$-antisymmetry we mean that the exchange of two quarks is the negative of (\ref{quarkstatistics}) with eigenvalue $-q$. The isospin states $\phi_S\left(\frac{1}{2},\frac{1}{2}\right)$ and $\phi_A\left(\frac{1}{2},\frac{1}{2}\right)$ can be identified with the mixed symmetric and mixed anti-symmetric flavor states of the proton. Similarly $\phi_S\left(\frac{1}{2},\frac{-1}{2}\right)$ and $\phi_A\left(\frac{1}{2},\frac{-1}{2}\right)$ are those of the neutron. The equivalent flavor states for the remaining octet baryons are obtained via application of the coproduct of the second simple root $\Delta\hat{U}_{\pm}$. The full set of mixed symmetry and mixed anti-symmetry flavor states for the octet baryons are listed in appendix A.

We do not in this paper deform the spin space and thus the symmetric and anti-symmetric spin states are the classical ones
\begin{eqnarray}
\chi_S\left( \frac{1}{2},-\frac{1}{2}\right)  &=& \frac{-1}{\sqrt{6}}\left( 2\downarrow\downarrow\uparrow-\uparrow\downarrow\downarrow-\downarrow\uparrow\downarrow \right),\\
\chi_S\left( \frac{1}{2},\frac{1}{2}\right)  &=& \frac{1}{\sqrt{6}}\left(2\uparrow\uparrow\downarrow-\downarrow\uparrow\uparrow-\uparrow\downarrow\uparrow\right),\\
\chi_A\left( \frac{1}{2},\frac{-1}{2}\right)  &=&\frac{-1}{\sqrt{2}}\left(\downarrow\uparrow\downarrow -\uparrow\downarrow\downarrow \right) ,\\
\chi_A\left( \frac{1}{2},\frac{1}{2}\right)  &=&\frac{1}{\sqrt{2}}\left( \uparrow\downarrow\uparrow-\downarrow\uparrow\uparrow\right),
\end{eqnarray}
where we have again supressed the tensor products in order to simplify the notation.

\section{Spin-flavor proton and neutron states and their magnetic moments}\label{protonneutron}

Using the deformed flavor states and undeformed spin states of the previous section, the spin-flavor state of a spin-up proton can now be written as
\begin{eqnarray}
\nonumber\vert p \uparrow \rangle &=& \frac{1}{\sqrt{2}}\left[ \phi_S\left( \frac{1}{2},\frac{1}{2}\right)\chi_S\left( \frac{1}{2},\frac{1}{2}\right)+\phi_A\left( \frac{1}{2},\frac{1}{2}\right)\chi_A\left( \frac{1}{2},\frac{1}{2}\right)\right],\\
\nonumber&=&\frac{q^{-1/4}}{\sqrt{2[2]_q [3]_q}}\left( [2]_q q^{1/2}uud-duu-q^{-1/2}udu\right)\frac{1}{\sqrt{6}}\left(2\uparrow\uparrow\downarrow-\downarrow\uparrow\uparrow-\uparrow\downarrow\uparrow\right)\\
\nonumber &+&\frac{1}{\sqrt{2[2]_q}}\left( q^{-1/4}udu-q^{1/4}duu\right)\frac{1}{\sqrt{2}}\left( \uparrow\downarrow\uparrow-\downarrow\uparrow\uparrow\right).
\end{eqnarray}
Expanding one obtains
\begin{eqnarray}\label{singledef}
\nonumber\vert p \uparrow \rangle &=&\left[ 2[2]_qq^{1/2}u\uparrow u\uparrow d\downarrow-[2]_qq^{1/2}u\uparrow u\downarrow d\uparrow-[2]_qq^{1/2}u\downarrow u\uparrow d\uparrow-2d\uparrow u\uparrow u\downarrow\right.\\
\nonumber&+&\left.(1-\sqrt{3[3]_q})d\uparrow u\downarrow u\uparrow+ (1+\sqrt{3[3]_q})d\downarrow u\uparrow u\uparrow-2q^{-1/2} u\uparrow d\uparrow u\downarrow\right.\\
\nonumber&+&\left.(q^{-1/2}+q^{1/2}\sqrt{3[3]_q})u\uparrow d\downarrow u\uparrow+(q^{-1/2}-q^{1/2}\sqrt{3[3]_q})u\downarrow d\uparrow u\uparrow\right].
\end{eqnarray}
From this expression the magnetic moment of the proton may be calculated to be \footnote{Some of the explicit calculations are provided in appendix B.} 
\begin{eqnarray}
\nonumber\mu_p=\frac{q^{-1/2}}{6[2]_q[3]_q}\left[ (7q^{1/2}[2]_q[3]_q-3q^{-1/2}[2]_q+4\sqrt{3[3]_q})\mu_u\right.\\-\left. (q^{1/2}[2]_q[3]_q-3q^{-1/2}[2]_q+4\sqrt{3[3]_q})\mu_d\right]\label{protonmm},  
\end{eqnarray}
which we simply rewrite as
\begin{eqnarray}
\mu_p= A\mu_u+(1-A)\mu_d,
\end{eqnarray}
where
\begin{eqnarray}
A=\frac{q^{-1/2}}{6[2]_q[3]_q}(7q^{-/2}[2]_q[3]_q-3q^{-1/2}[2]_q+4\sqrt{3[3]_q}).
\end{eqnarray}
It is readily checked that in the limit $q\rightarrow 1$, that $A=4/3$ as required.

A similar calculation for the magnetic moment of the neutron gives
\begin{eqnarray}
\nonumber\mu_n=\frac{q^{1/2}}{6[2]_q[3]_q}\left[ (7q^{-1/2}[2]_q[3]_q-3q^{1/2}[2]_q+4\sqrt{3[3]_q})\mu_u\right.\\+\left. (-q^{-1/2}[2]_q[3]_q+3q^{1/2}[2]_q-4\sqrt{3[3]_q})\mu_d\right]\label{neutronmm},  
\end{eqnarray}
which we rewrite as
\begin{eqnarray}
\mu_n= B\mu_u+(1-B)\mu_d.
\end{eqnarray}
with
\begin{eqnarray}
B&=&\frac{q^{1/2}}{6[2]_q[3]_q}(-q^{-1/2}[2]_q[3]_q+ 3q^{1/2}[2]_q-4\sqrt{3[3]_q}).
\end{eqnarray}

One may likewise calculate the magnetic moments of the $\Sigma^+,\Sigma^-,\Xi^0,\Xi^-$ baryons,
\begin{eqnarray}
\mu_{\Sigma^+}&=&A\mu_u+(1-A)\mu_s,\\
\mu_{\Sigma^-}&=&A\mu_d+(1-A)\mu_s,\\
\mu_{\Xi^0}&=&B\mu_u+(1-B)\mu_s,\\
\mu_{\Xi^-}&=&B\mu_d+(1-B)\mu_s.
\end{eqnarray}

\section{Masses of the up and down quarks}\label{quarkmasses}

Equations (\ref{protonmm}) and (\ref{neutronmm}) can be used to obtain expressions for $\mu_u$ and $\mu_d$ as functions of $\mu_p$, $\mu_n$, and $q$. 
\begin{eqnarray}
\mu_u &=& \frac{1}{A-B}\left[ (1-B)\mu_p+(A-1)\mu_n\right]\label{upmm},\\
\mu_n &=& \frac{1}{A-B}\left[ -B\mu_p+A\mu_n\right]\label{downmm} .
\end{eqnarray}
Using the experimentally observed values for  $\mu_p$ and $\mu_n$ we can then determine the masses of the up and down quarks as a function of the deformation parameter. Insisting that both magnetic moments and masses must be real limits means that the deformation parameter $q$ can only take on real values. Figure 1 shows the masses of the up (solid red line) and down (dashed blue line) quarks as a function of the (real) deformation parameter $q$.
\begin{figure}[ht]\label{quarkmasses2}
\centering
   \includegraphics[scale=0.3]{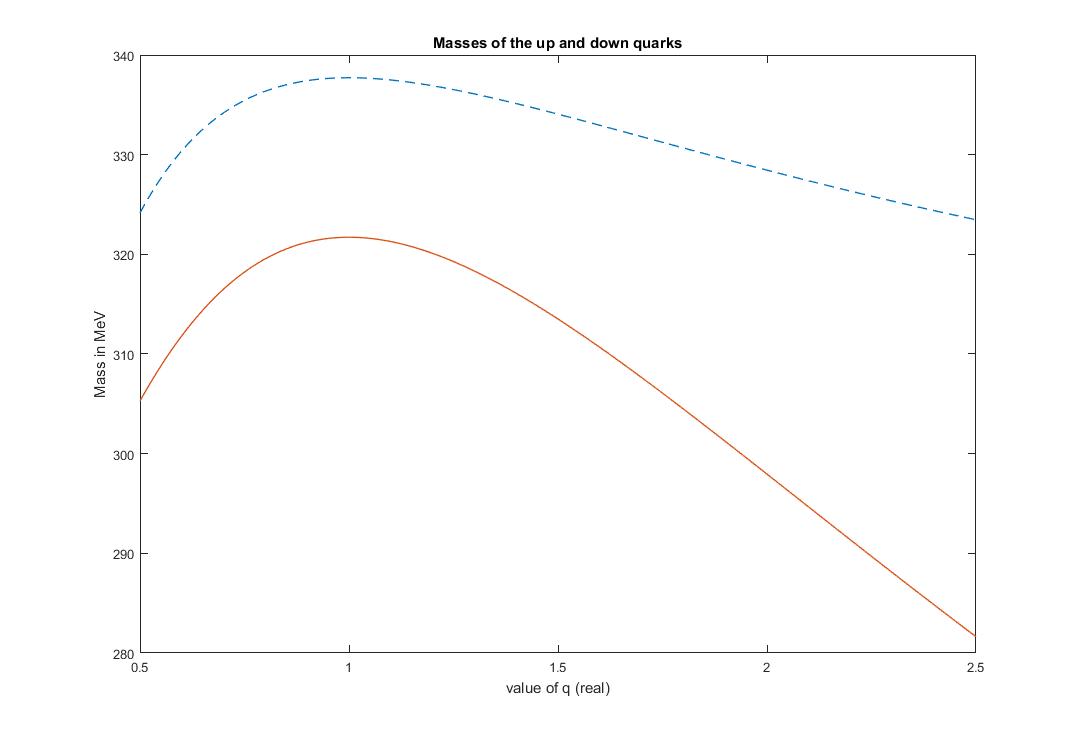}
\caption{Masses of the up (solid red line) and down (dashed blue line) quarks as a function of the (real) deformation parameter $q$ for the case where the flavor space is deformed. The undeformed case $q=1$ corresponds to a maximum for both masses.}
\end{figure}
\section{The $\Lambda$ flavor state and the magnetic moment of the strange quark}\label{lambdastate}

The construction of the $\Lambda$ octet baryon requires some care. Starting with a general $uds$ quark state (6 terms) the following conditions are imposed:
\begin{enumerate}
\item $q$-(anti)Symmetric under the interchange of the first two quarks,
\item Orthogonal to the (anti)symmetric $\Sigma^0$ flavor state,
\item Vanishes under $\hat{T}_{\pm}$,
\item Orthogonal to the totally q-antisymmetric flavor singlet.
\end{enumerate}
Imposing these conditions gives the mixed $q$-symmetric and mixed $q$-antisymmetric flavor states
\begin{eqnarray}
\nonumber\vert \Lambda\rangle_S&=&\frac{1}{[2]_q}\left( q^{1/2}sud+usd-sdu-q^{-1/2}dsu\right) ,\\
\nonumber\vert \Lambda\rangle_A&=&\frac{q}{[2]_q\sqrt{[3]_q}}\left( -q^{-3/2}[2]_qdus+q^{-1}[2]_q uds+usd-q^{-1/2}dsu-q^{-1/2}sud+q^{-1}sdu\right). 
\end{eqnarray}
The magnetic moment of the $\Lambda$ can be calculated by combining the above flavor states with the spin states. Following a tedious calculation one obtains
\begin{eqnarray}
\mu_{\Lambda}=C\mu_u+ D\mu_d+ E\mu_s,
\end{eqnarray}
where
\begin{eqnarray}
\nonumber C&=&\frac{1}{12[2]_q^2[3]_q}\left( 4(q^{1/2}-q^{-1/2})[2]_q[3]_q+6q^{1/2}[2]_q+2q^{-1/2}[2]_q[3]_q-8q\sqrt{3[3]_q}\right) ,\\
\nonumber D&=&\frac{1}{12[2]_q^2[3]_q}\left(-4(q^{1/2}-q^{-1/2})[2]_q[3]_q+6q^{3/2}[2]_q+2q^{1/2}[2]_q[3]_q-8\sqrt{3[3]_q}\right) ,\\
\nonumber E&=&\frac{1}{12[2]_q^2[3]_q}\left(4[2]_q^2[3]+6q^{-1/2}[2]^3+8q^{1/2}[2]_q\sqrt{3[3]_q}\right) 
\end{eqnarray}
Despite the lack of elegance of this expression, it is easy to check that each term exhibits the correct behavior in the $q\rightarrow 1$ limit, namely $C=D=0,\;E=1$. Furthermore, $C+D+E=1$ independent of $q$. Rearranging, we obtain an expression for the magnetic moment of the strange quark
\begin{eqnarray}\label{strangemm}
\mu_s=\frac{1}{E}(\mu_{\Lambda}-C\mu_u-D\mu_d)
\end{eqnarray}

\section{Comparison with experimental data}\label{analysis}

Substituting the experimentally observed values for $\mu_p$, $\mu_n$, and $\mu_{\Lambda}$ into equations (\ref{upmm}), (\ref{downmm}), and (\ref{strangemm}) gives the magnetic moments of the three light quarks as functions of $q$. This in turn allows us to express the magnetic moments of all the octet baryons in terms of $q$. Since we have used the proton, neutron and Lambda magnetic moments as input, and since the is no experimental data for the magnetic moment of the $\Sigma^0$ baryon, we calculate the least square error for the magnetic moments of the four remaining octet baryons ($\Sigma^+, \;\Sigma^-,\;\Xi^0,\;\Xi^-$) as a function of $q$. Plotting this function we are then able to find the values of $q$ which correspond to the small least square error. This plot is shown in figure 2 below.
\begin{figure}[ht]\label{lse}
\centering
   \includegraphics[scale=0.38]{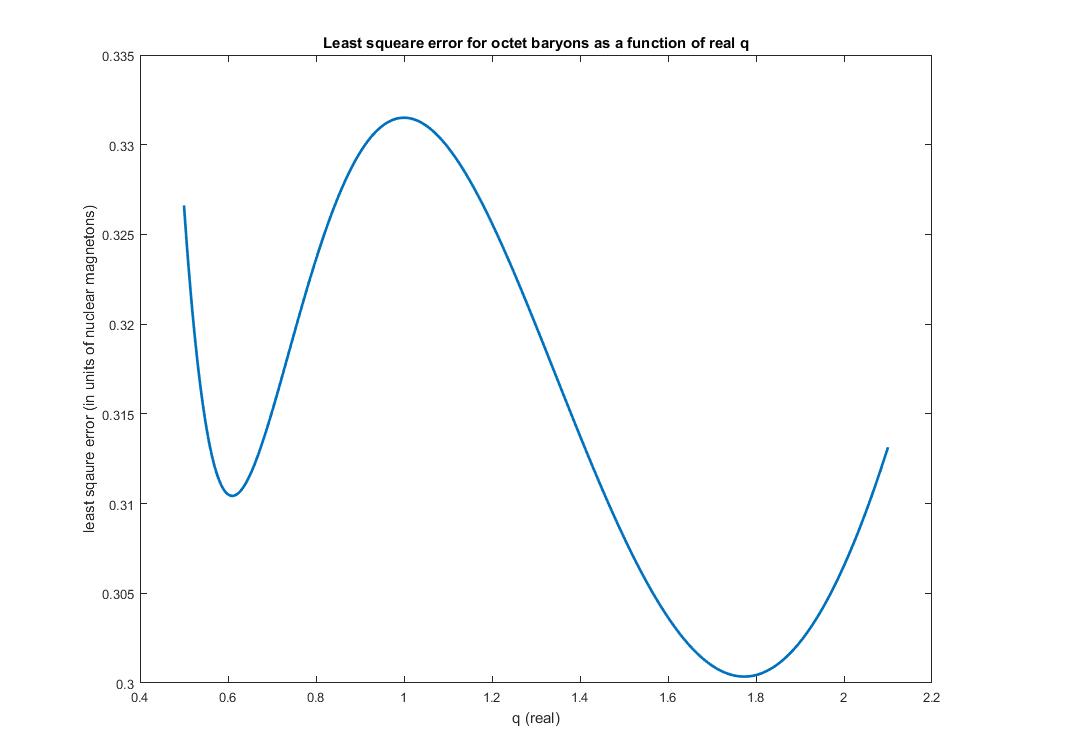}
\caption{Least square error of the octet baryon magnetic moments for the case where the flavor space is deformed.}
\end{figure}
The graph shows that the least square error has a local maximum for the undeformed $q=1$ case. A small deformation results in a better fit to experimental data with a local minimum when $q=0.610$ and a global minimum when $q=1.774$. When $q=1.774$, the least square error in units of nuclear magnetons is 0.3004, compared to 0.3315 for the undeformed $q=1$ case. This represents a roughly $10\%$ reduction in error. These main results, as well as the associated masses of the up and down quarks at these values of $q$ are summarized in Table 1.

\begin{table}[hb]\label{table1}
\caption{Table showing the up and down quark masses for values of $q$ corresponding to a local minimum least square error.}
\centering
\begin{tabular}{c c c c}
\hline\hline
$q$ & $m_u$ (MeV) & $m_d$ (MeV) & LSE ($\mu_N$)\\
\hline
1 & 338 &322 & 0.3315\\
1.774 & 331 & 305 & 0.3004\\
0.610 & 331 & 312 & 0.3104\\
\hline 
\end{tabular}
\label{table:results}
\end{table}
\newpage
\section{Discussion}

In this paper we have investigated the effect of using a $q$-deformed flavor symmetry on the magnetic moments of octet baryons. Our work was motivated by the exceptionally accurate baryon mass sum rules that arise from deforming flavor symmetry \cite{gavrilik1997quantum,gavrilik2001quantum,Gresnigt2016baryons}. 

Using the experimentally observed values for the magnetic moments of the proton, neutron and $\Lambda$ baryons as input, we calculated the magnetic moments of the up, down, and strange quarks, and by extension those of the remaining octet baryons as functions of the deformation parameter $q$. For certain real values of $q$, the deformed magnetic moments fit the experimental data better. In particular, the value $q=1.774$ gives an approximately $10\%$ decrease in the least square error compared to the undeformed case corresponding to $q=1$. For this value of $q$ the quark masses of the up and down quark are calculated to be 331 MeV and 305 MeV respectively. 

Deforming only the flavor space means the classical notion of a fermion is lost. This is because the flavor states are no longer eigenstates of the permutation operator but rather of the braid operator with eigenvalues not always equal to $\pm1$. To restore the classical notion of a fermion one could consider additional deformations in the spatial, spin, and color spaces. Given the encouraging results of the present investigation, a more complete study that looks at multiple deformations and includes the decuplet baryons seems a logical next step.

Quantum groups are the result of one-parameter deformation in the universal enveloping algebra (a Hopf algebra). These are just one type of algebraic deformation that can be considered \cite{gerstenhaber1964,nijenhuis1967}. Lie-type deformations (those that deform a Lie algebra) have proven very useful in generalizing spacetime symmetries \cite{mendes1994,ahluwalia2008ppa,Chryssomalakos2004,gresnigt2007sph}. $q$-Deformation on the other hand seem to have particular applications in generalized descriptions of internal and gauge symmetries \cite{Finkelstein2012,Finkelstein2005,finkelstein2007elementary}. Considered together, Lie algebra and Hopf algebra deformations offer a consistent framework within which to develop physics in the 21st century \cite{gresnigt2015electroweak,Sternheimer2014,Flato1982,bonneau2003topological}.
 
\section*{Acknowledgments}
This work is supported in part by XJTLU research grant RDF-14-03-13 and Natural Science Foundation of China grant RR0116.

\section*{Appendix A: Deformed octet flavor states}\label{deformedflavorstates}

The $q$-octuplet states are: 
\begin{eqnarray}
\nonumber\phi(p)_S&=&\left|\frac{1}{2},1\right\rangle_S=\frac{q^{\frac{-1}{4}}}{\sqrt{[2]_q[3]_q}}[[2]_qq^{\frac{1}{2}}uud-duu-q^{\frac{-1}{2}}udu], \\
\nonumber\phi(p)_A&=&\left|\frac{1}{2},1\right\rangle_A=\frac{1}{\sqrt{[2]_q}}[q^{\frac{-1}{4}}duu-q^{\frac{1}{4}}udu], \\
\nonumber\phi(n)_S&=&\left|\frac{-1}{2},1\right\rangle_S=\frac{-q^{\frac{1}{4}}}{\sqrt{[2]_q[3]_q}}[[2]_qq^{\frac{-1}{2}}ddu-udd-q^{\frac{1}{2}}dud], \\
\nonumber\phi(n)_A&=&\left|\frac{-1}{2},1\right\rangle_A=\frac{1}{\sqrt{[2]_q}}[q^{\frac{-1}{4}}dud-q^{\frac{1}{4}}udd], \\
\nonumber\phi(\Sigma^+)_S&=&\left|1,0\right\rangle_S=\frac{q^{\frac{-1}{4}}}{\sqrt{[2]_q[3]_q}}[[2]_qq^{\frac{1}{2}}uus-suu-q^{\frac{-1}{2}}usu], \\
\nonumber\phi(\Sigma^+)_A&=&\left|1,0\right\rangle_A=\frac{1}{\sqrt{[2]_q}}[q^{\frac{-1}{4}}suu-q^{\frac{1}{4}}usu], \\
\nonumber\phi(\Sigma^-)_S&=&\left|-1,0\right\rangle_S=\frac{q^{\frac{-1}{4}}}{\sqrt{[2]_q[3]_q}}[[2]_qq^{\frac{1}{2}}dds-sdd-q^{\frac{-1}{2}}dsd], \\
\nonumber\phi(\Sigma^-)_A&=&\left|-1,0\right\rangle_A=\frac{1}{\sqrt{[2]_q}}[q^{\frac{-1}{4}}sdd-q^{\frac{1}{4}}dsd], \\
\nonumber\phi(\Sigma^0)_S&=&\left|0,0\right\rangle_S=\frac{-q^{\frac{1}{4}}}{\sqrt{[2]_q[3]_q}}[[2]_qq^{\frac{-3}{4}}dsu+[2]_qq^{\frac{-1}{4}}sdu-q^{\frac{1}{4}}usd-q^{\frac{1}{4}}dus-q^{\frac{-1}{4}}uds-q^{\frac{3}{4}}sud], \\
\nonumber\phi(\Sigma^0)_A&=&\left|0,0\right\rangle_A=\frac{1}{[2]_q}[q^{\frac{-1}{2}}dus+sud-uds-q^{\frac{1}{2}}usd], \\
\nonumber\phi(\Xi^-)_S&=&\left|\frac{-1}{2},-1\right\rangle_S=\frac{-q^{\frac{1}{4}}}{\sqrt{[2]_q[3]_q}}[[2]_qq^{\frac{-1}{2}}ssd-dss-q^{\frac{1}{2}}sds], \\
\nonumber\phi(\Xi^-)_A&=&\left|\frac{-1}{2},-1\right\rangle_A=\frac{1}{\sqrt{[2]_q}}[q^{\frac{-1}{4}}sds-q^{\frac{1}{4}}dss], \\
\nonumber\phi(\Xi^0)_S&=&\left|\frac{1}{2},-1\right\rangle_S=\frac{-q^{\frac{1}{4}}}{\sqrt{[2]_q[3]_q}}[[2]_qq^{\frac{-1}{2}}ssu-uss-q^{\frac{1}{2}}sus], \\
\nonumber\phi(\Xi^0)_A&=&\left|\frac{1}{2},-1\right\rangle_A=\frac{1}{\sqrt{[2]_q}}[q^{\frac{-1}{4}}sus-q^{\frac{1}{4}}uss], \\
\nonumber\phi(\Lambda^0)_S&=&\frac{1}{[2]_q}[q^{\frac{1}{2}}sud+usd-sdu-q^{\frac{-1}{2}}dsu], \\
\nonumber\phi(\Lambda^0)_A&=&\frac{q}{[2]_q\sqrt{[3]_q}}[-q^{\frac{-3}{2}}[2]_qdus+q^{-1}[2]_quds+usd-q^{\frac{-1}{2}}dsu-q^{\frac{-1}{2}}sud+q^{-1}sdu].
\end{eqnarray}

\section*{Appendix B: Calculation of magnetic moment of proton}

\begin{eqnarray}
\nonumber\vert p \uparrow \rangle &=&\left[ 2[2]_qq^{1/2}u\uparrow u\uparrow d\downarrow-[2]_qq^{1/2}u\uparrow u\downarrow d\uparrow-[2]_qq^{1/2}u\downarrow u\uparrow d\uparrow-2d\uparrow u\uparrow u\downarrow\right.\\
\nonumber&+&\left.(1-\sqrt{3[3]_q})d\uparrow u\downarrow u\uparrow+ (1+\sqrt{3[3]_q})d\downarrow u\uparrow u\uparrow-2q^{-1/2} u\uparrow d\uparrow u\downarrow\right.\\
\nonumber&+&\left.(q^{-1/2}+q^{1/2}\sqrt{3[3]_q})u\uparrow d\downarrow u\uparrow+(q^{-1/2}-q^{1/2}\sqrt{3[3]_q})u\downarrow d\uparrow u\uparrow\right] 
\end{eqnarray}
Calculating the magnetic moments the $(2\mu_u-\mu_d)$ factor is
\begin{eqnarray}
&(&2[2]_q^2q^{1/2})^2+(1+\sqrt{3[3]_q})^2+(q^{-1/2}+q^{1/2}\sqrt{3[3]_q})^2\\
&=&4q([3]_q+1)+1+4\sqrt{3[3]_q}+3[3]_q+q^{-1}+3q[3]_q\\
&=&7q[3]_q+4q+3[3]_q+1+q^{-1}+4\sqrt{3[3]_q}\\
&=&7(q+1)[3]_q-(1+q^{-1})3+4\sqrt{3[3]_q}\\
&=&7q^{1/2}[2]_q[3]_q-3q^{-1/2}[2]_q+4\sqrt{3[3]_q}
\end{eqnarray}

\bibliography{NielsReferences}  
\bibliographystyle{unsrt}  

\end{document}